\documentclass[usenatbib,times]{mn2e}
\usepackage{graphicx}
\usepackage{epsf}

\def\xte{{\it RXTE}}

\def\H0{{\rm ~km~s^{-1}~Mpc^{-1}}}

\begin{document}

\title[Rapid variability around black holes]{Fast variability as a probe of the smallest regions around accreting black holes}

\author[Axelsson et al.]{Magnus Axelsson,$^{1,2}$\thanks{email: magnusa@astro.su.se}
Linnea Hjalmarsdotter$^{3}$ and Chris Done$^{4}$\\
$^{1}$Oskar Klein Center for CosmoParticle Physics, Department of Physics, Stockholm University, SE-106 91 Stockholm, Sweden\\
$^{2}$Department of Astronomy, Stockholm University, SE-106 91 Stockholm, Sweden\\
$^{3}$Sternberg Astronomical Institute, Moscow State University, Universitetskij pr. 13, 119899 Moscow, Russia\\
$^{4}$Department of Physics, Durham University, South Road, Durham DH1 3LE, UK\\
}

\date{Accepted --. Received --; in original form --}

\pagerange{\pageref{firstpage}--\pageref{lastpage}} \pubyear{2002}

\maketitle

\begin{abstract}

We extract the spectra of the fastest variability (above 10 Hz) from
the black hole XTE J1550-564 during a transition from hard to soft
state on the rise to outburst. We confirm previous results that the
rapid variability contains no significant disc component despite this
being strongly present in the total spectrum of the softer
observations.  We model ionised reflection significantly better than
previous work, and show that this is also suppressed in the rapid
variability spectrum compared to the total emission. This is
consistent with the fast variability having its origin in a hot inner
flow close to the black hole rather than in the accretion disc or in a
corona above it. However, the rapid variability spectrum is not simply
the same as the total Comptonised emission. It is always significantly
harder, by an amount which increases as the spectrum softens during
the outburst. This adds to evidence from time lags that the
Comptonisation region is inhomogeneous, with harder spectra produced
closest to the black hole, the same region which produces the fastest
variability.

\end{abstract}

\section{Introduction}

Spectral modelling is a powerful tool in understanding the physics of astrophysical sources including accreting black holes. By assigning observed/modelled spectral components to physical regions of the accretion flow we can learn about its geometry and how it changes as a response to changes in accretion rate.

The most popular physically motivated model for the X-ray spectra of accreting black holes is the combination of an accretion disc modelled as a multi-temperature blackbody  and an inner hot flow and/or a corona above the disc, with hot electrons up-scattering the disc photons giving rise to a Comptonized component. Many spectra of black holes, especially in the soft/high to very high states, also require the presence of reflection of the Comptonized photons off some cooler material, presumably the accretion disc, and various lines, mostly from reflection and absorption from iron or other heavy elements. Such models of disc+Comptonization+reflection(+lines), modified by line-of-sight absorption, have been shown to rather successfully describe the spectra from black holes in a variety of spectral states at a range of accretion rates \citep[for a review of spectra of black holes see][]{done07}.

In reality, however, inadequate models for the disc and reflection
combined with poorly known parameters for Comptonization often make
these components rather hard to separate, and spectral modelling
therefore inherently suffers from degeneracy. In addition, more
realistic modelling of the increasing amount of high quality data
often require more complexity than the 'basic' components. Especially
for more complex models, a unique decomposition of the emission
components may not be possible with spectral analysis alone, as
there are many combinations which match a given observed spectrum.

Information about the accretion geometry may also be gleaned from
analysis of temporal variability. The X-ray emission from black hole
binaries is variable on timescales from milliseconds to days. While
the slower variations can be directly coupled to changes in the
accretion rate, the source of rapid variability is more
uncertain. Like the case with radiation spectra, there are likely
several components to the variability. The broad band variability is
commonly studied through the power density spectrum (PDS). For black
holes in the low/hard to intermediate (including so called very high)
states the PDS can typically be well-described by a combination of
Lorentzian components (Belloni, Psaltis \& van der Klis 2002)
in the range $0.01-100$
Hz, while the PDS during transitions into soft, strongly
disc-dominated states is more variable and not easily decomposed. In
addition to broad band variability, many sources display
quasi-periodic oscillations (QPOs), varying both with the spectral
state of the source as well as the energy band being studied.

In this paper, we use the technique of frequency resolved spectroscopy
(Revnivtsev, Gilfanov \& Churazov 1999) to derive the (rms) energy spectrum of the
variability. The significance of the rms energy spectrum is that it
describes the energy spectrum of the process responsible for the
variability, assuming the variability is generated by fluctuations in
the normalization of a separate spectral component. The energy
spectrum of the fast variability can then be directly compared to the
continuum energy spectrum and help to constrain the observed
components (e.g., Churazov, Gilfanov \& Revnivtsev 2001; Revnivtsev \& Gilfanov 2006; Sobolewska \& {\.Z}ycki 2006).

\section{XTE J1550-564}

The transient black hole binary XTE~J1550-564 was discovered on 1998 September 7 by the {\xte} All-Sky Monitor \citep[ASM;][]{smi98} and the {\it CGRO} Burst and Transient Source Experiment \citep[BATSE;][]{wil98}. This source has showed four subsequent outbursts since its discovery, but the first outburst remains the brightest and best covered by {\xte} pointing observations. The data span nearly three decades in X-ray luminosity, capturing the source in each of the canonical states typically associated with black hole binaries. The broad band spectrum of XTE~J1550-564 has previously been modelled by several authors \citep[e.g.,][]{sob99,tom01,gd03} including the data used here.

The variability of XTE~J1550-564 is similar to that of other black
hole binaries, and changes with the accretion rate. The power density
spectrum in the low/hard state is characterized by strong aperiodic
noise. The total fractional variability is high, up to $\sim30$\%. The
continuum of the power spectrum can usually be approximated as a
broken power law with a flat power spectrum (in $fP_f$) 
between a low and high frequency break, decreases at 
frequencies above and below these. The PDS evolves
during the rising part of an outburst, with the low frequency break 
moving to higher frequencies, with much less change at high
frequencies (Psaltis, Belloni \& van der Klis 1999; 
Gierlinski, Nikolajuk \& Czerny 2008). 
As the source enters a high/soft state, the power spectrum
changes drastically, 
losing the low frequency break
\citep[see for example][]{rem99}. The amplitude of variability also
decreases dramatically, but this is due to the increasing dominance of
the stable disc component in the spectrum (Churazov et al. 2001). 

XTE~J1550-564 also exhibits transient high-frequency quasi-periodic
oscillations (HFQPOs) as well as a stronger and more common
low-frequency QPO \citep[LFQPO;][]{cui99,rem02}. The latter is seen to
vary in frequency in response to changes in the radiation spectrum,
coupled to variations in the accretion rate and geometry. In some observations a
second low frequency feature appears with harmonic content. Such a
feature has also been seen in other black hole transient systems,
e.g. GX~339-4 \citep{bel05}.

\section{Data analysis}

In this study we use archival data of XTE~J1550-564 from the Proportional Counter Array \citep[PCA;][]{jah96} instrument onboard the {\xte} satellite. The observations are made 
in the time period 1998-09-09 to 1998-09-16 (MJD 51065 to 51072) and cover the rise of the strong initial 
flare of the 1998 outburst \citep[see][for more details on this outburst]{smi98}. We extracted Standard2 spectra for each observation applying standard selection criteria. A 
systematic error of 1 per cent was added to each bin in the spectra. 
This dominates over the statistical uncertainties, but we still use
the standard $\Delta \chi^2=2.7$ to determine error ranges. 
The energy band used in the spectral modelling was 3--20 keV.

\subsection{Spectral model}

The data were fit in \textsc{xspec} using a spectral model comprising
a disc blackbody, thermal Comptonization and Compton reflection.  The
disc component was modelled using {\sc diskbb}, where the main
parameter is the maximum temperature $T_{\rm bb}$. As our data only
cover energies above 3 keV, far above the peak of the disc component,
the disc temperature is poorly constrained so while we allow this to
be free in the fits, we have to freeze it at the best-fit value when
calculating the error ranges for the other parameters.

\begin{figure*}
\includegraphics[width=\textwidth,angle=0]{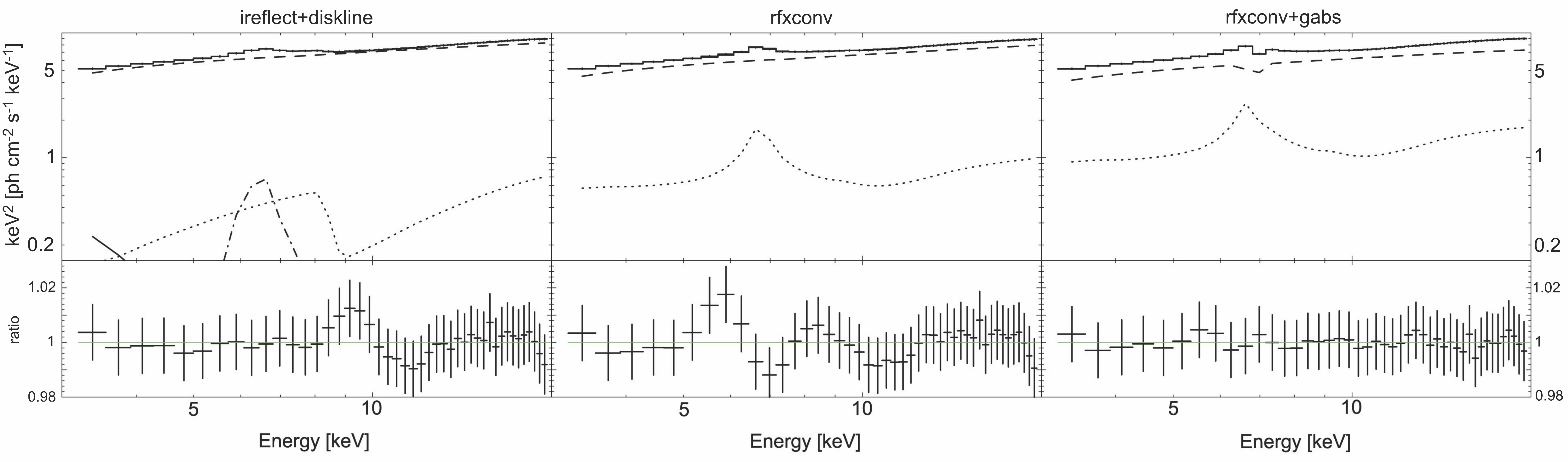}
\caption{Fit to PCA data for spectrum 30188-06-01-00 with residuals, showing the results when using different models for the reflection. The dashed and dotted lines show the contribution from thermal Comptonization and Compton reflection, respectively. The models shown here are {\sc ireflect+diskline} (left panel; dash-dot line shows the line component), {\sc rfxconv} (middle) and {\sc rfxconv+gabs} (right panel). Although all fits are statistically very good ($\chi^2$/dof $<0.5$), the residuals clearly show that the region around the K$\alpha$ iron line is not always well modelled. In this paper we use {\sc rfxconv+gabs}, which provides the best model to the data.}
\label{pcafits}
\end{figure*}

For Comptonization we use the model {\sc nthcomp} \citep{zdz96,zyc99},
parametrized by the asymptotic spectral index $\Gamma$, the seed
photon temperature $T_0$ and the electron temperature $kT_{\rm e}$. We
set $T_0 = T_{\rm bb}$ as can be expected if the seed photons are the
photons from the disc. As it can be expected that the seed photons
come from the inner edge of the disc, we choose a blackbody (rather
than a disc spectrum) as the input type.

For Compton reflection we use {\sc rfxconv}. This is a convolution
model that combines the reflected spectra of Ross \& Fabian (2005)
from a constant density ionized disc, with the
\textsc{ireflect} convolution variant of the \textsc{pexriv} Compton
reflection code by Magdziarski \& Zdziarski (1995). The table models
calculate the reflected spectrum and iron line from an X-ray illuminated slab
including Compton up- and down-scattering below 12 keV but use a fixed
spectral form (exponentially cutoff power law). By contrast, the
\textsc{ireflect} model is a full convolution model for any continuum
shape but neglects Compton scattering below 12 keV, has very
approximate ionisation balance and does not include the self-consistent iron line emission. 
The combination of these models into the new
{\sc rfxconv} model thus allows us to model a physically much more
correct reflection spectrum for the whole PCA energy range including a
self-consistently calculated iron line. The \textsc{rfxconv} model was
demonstrated and briefly described in \citet{koh11} and is an updated
version of the model first described in \citet{dg06}. The main
parameters are the relative amplitude of the reflected component, $R$,
the inclination and the ionization parameter of the reprocessing
matter, $\xi$.

We include photoelectric absorption, keeping the value of the column
density frozen at $N_{\rm H}=0.6\times 10^{22}$ cm$^{-2}$ following
the modelling in \citet{gd03}. We also find that in all the data sets
our fits are significantly improved by the addition of a narrow
absorption line at $\sim6.8$ keV. Observations of similar strong
absorption lines have been reported in several high inclination
Galactic binary systems and are believed to represent iron K resonance
at 6.7 and 7.0 keV from outflowing material (e.g., Ueda et al. 1998;
Kotani et al. 2000; Lee et al. 2002; Ueda et al. 2004). Due to the
limited resolution of the PCA data, we model the absorption line as a
gaussian of fixed (0.1~keV) intrinsic width. With the use of the
\textsc{rfxconv} model plus the absorption line our fits show
virtually no residuals in the complex iron line
region. Fig.~\ref{pcafits} shows an example fit to one of the
studied spectra using \textsc{rfxconv+gabs} in comparison to the same
data fit with only \textsc{rfxconv} and with
\textsc{ireflect+diskline}.

\subsection{Fourier Spectroscopy}
\label{fourspec}
To compare the total spectrum to that of the rapid variability, we also performed frequency-resolved spectroscopy \citep{rev99,rev01}. Following the approach of \citet{rev99},
we extracted a light curve for each available channel and constructed a power density spectrum. By integrating the PDS over the frequency range of interest, we 
determined the relative contribution for each channel, and combined these to make an energy spectrum of the rapid variability. Throughout our observations, a strong QPO 
was present at $\leq 4$ Hz, and a feature of harmonic content was seen in many observations as well, as reported earlier by \citet{cui99} and \citet{rem02}, among others. In order to avoid contamination 
from either the QPO or the harmonic, we chose the frequency range 10--50 Hz. As in the case of the Standard2 spectra, we added a 1 per cent systematic error. 

Using the same technique, we also extracted the energy spectrum of the QPO. To obtain its contribution in each energy band, we fit the QPO feature in 
the PDS using a Lorentzian function, and integrate over this component \citep[see also][]{sob06}. This gives a more accurate measure of its contribution than simply integrating the PDS over the energy range where the QPO dominates.

\section{Results}

\subsection{The total PCA spectra}

In the first step, we fit the PCA Standard2 spectra with our model
{\textsc{tabs$\times$gabs$\times$(diskbb+nthcomp+} {\sc
nthcomp$\times$rfxconv})}. For all spectra we find a very good
fit, with  $\chi^2$/dof $<0.3$ showing that the systematic error of
1\% dominates the statistical errors. 
The residuals clearly show that our improved
model for the reflection matches the data very well. In particular, we
note that our use of \textsc{rfxconv} plus the absorption line enables
us to make an good fit in the difficult region around the iron line,
something which has proven difficult using earlier available models
(e.g. Done \& Kubota 2006).

Despite the limited bandwidth at low energies of the PCA, our model
gives very reasonable, although not very well constrained, disc
temperatures in the range of 0.5-0.7 keV, typical for intermediate/very
high states. We note that the reason we are able to constrain the 
disc temperature at all is that we are using it as the seed photon 
temperature. Changing the disc temperature thereby affects also 
the Comptonized component of the spectrum.
The asymptotic spectral index $\Gamma$ increases from 1.8
in the hardest states to 2.4 in the softest. The electron temperature
in our fits, 9-15 keV, is significantly lower than expected for 
the hard spectral states, where more typical temperatures are around
100~keV (e.g. Ibragimov et al. 2005; Torii et al. 2011). However,
complex spectra with lower rollover temperatures are seen during
transitions \citep{gd03}. Clearly, data extending to  higher
energies are required to constrain the electron temperature (as well
as any possible existence of non-thermal Comptonization, which is not
considered in our model.

Reflection is significantly detected in all spectra, with solid angle
between 0.4 and 0.7, increasing with $\Gamma$ as expected as the disc
reaches further into the hot flow in the softer states (Revnivtsev et al. 2001).  
The ionization parameter of the reflector
is rather high, around $\log\xi=2.8$, and subsequently the peak of the
iron line complex falls around 6.7 keV. The absorption line energy
varies between 6.8 and 7.0 keV, consistent with resonant iron K
resonance energies for this low-resolution data.

\subsection{The spectrum of the variability}

\begin{figure}
\includegraphics[width=12cm,angle=270]{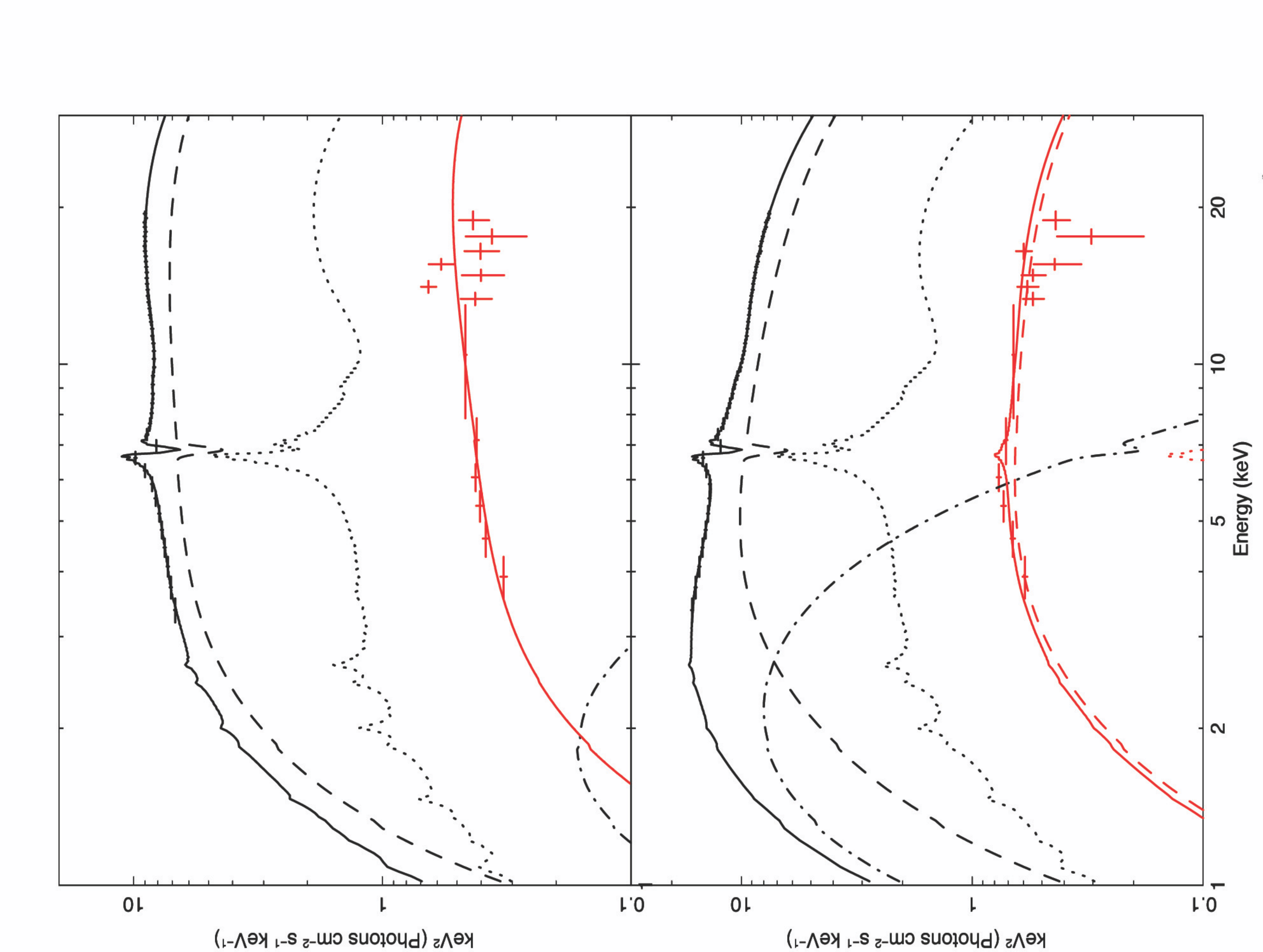}
\caption{Example spectra (data and model components) from XTE~J1550-564. The solid black spectrum shows a fit to the total continuum 3-20 keV PCA spectrum. The black dashed lines represent the Comptonized component, the dotted line the reflection and the dot-dashed line the disc blackbody, only required in the softer states (lower panel). The lower-normalization red spectrum shows a fit to the spectrum of the fast variability. This spectrum requires less reflection and does not require any thermal (blackbody) component.}
\label{fitex}
\end{figure}

We now add the spectra of the rapid variability to the fit. We
initially keep all parameters for both spectra frozen to the best-fit
values found for the PCA data alone, with the exception of the
normalizations of the disc blackbody and the Comptonized component as
well as the amplitude of reflection, which we allow to vary between
the continuum and variability spectra. In the harder observations,
this allows for an adequate fit also to the rms spectrum. However, as
the spectrum evolves and become softer, it is no longer possible to
find a good fit. We therefore allow also the asymptotic spectral index
$\Gamma$ to be different in the rms spectrum. This allows us to find
good fits to all datasets. Table~\ref{datatable} shows the resulting
parameter values. A decomposition of the fits to two of the datasets,
ObsIDs 30188-06-01-02 and 30188-06-09-00, showing the spectral
components is plotted in Fig.~\ref{fitex}.

\begin{table*}
\tabcolsep=0.12cm
\begin{tabular}{l l l l l l l l l l l}
ObsID & $T_{\rm bb}$ & $N_{\rm bb}$ & $\Gamma$ & $kT_{\rm e}$ &  $R$ & $E_{\rm line}$ & EW & $\log\xi$ & $\chi^2/dof$& Flux \\
& {\scriptsize (keV)} & & & {\scriptsize (keV)} & & {\scriptsize (keV)} & {(\scriptsize $\times 10^{-2}$ keV)} & & & {\scriptsize erg cm$^{-2}$ s$^{-1}$}\\
\hline
30188-06-01-00  & $0.56^{+0.31}_{-0.26}$ & $200^{+570}_{-200}$ & $1.77^{+0.03}_{-0.10}$ & $14.8^{+7.3}_{-3.0}$ & $0.44^{+0.12}_{-0.11}$ & $6.84^{+0.11}_{-0.11}$ & $9.5^{+4.0}_{-2.7}$ & $2.80^{+0.03}_{-0.09}$  & 2.3/36 & $4.4\times10^{-8}$ \\ [1ex]
rms  & & & $1.69^{+0.03}_{-0.03}$ & & $0.07^{+0.09}_{-0.07}$ & & & & 17.4/54 & \\ [1ex]

30188-06-01-01 & $0.55^{+0.04}_{-0.04}$ & $260^{+590}_{-260}$ & $1.80^{+0.03}_{-0.04}$ & $15.1^{+5.8}_{-3.0}$ & $0.43^{+0.14}_{-0.11}$ & $6.85^{+0.09}_{-0.09}$ & $9.2^{+3.5}_{-3.6}$ & $2.80^{+0.67}_{-0.09}$ & 7.0/36 & $4.0\times10^{-8}$ \\ [1ex]
rms & & &$1.66^{+0.09}_{-0.08}$ & & $0.06^{+0.25}_{-0.06}$ & & & & 50.8/54 & \\ [1ex]

30188-06-01-02 & $0.55^{+0.28}_{-0.08}$ & $410^{+1100}_{-410}$ & $1.89^{+0.02}_{-0.02}$ & $10.6^{+2.4}_{-1.5}$ & $0.56^{+0.15}_{-0.14}$ & $6.83^{+0.10}_{-0.09}$ & $11^{+5.0}_{-4.7}$ & $2.78^{+0.60}_{-0.09}$ & 4.5/36 & $4.3\times10^{-8}$  \\ [1ex]
rms & & & $1.78^{+0.03}_{-0.03}$ & & $0.0^{+0.07}_{-0.0}$ & & & & 35.2/54& \\ [1ex]

30188-06-01-03 & $0.53^{+0.25}_{-0.09}$ & $970^{+1100}_{-970}$ & $1.94^{+0.03}_{-0.09}$ & $10.9^{+2.3}_{-1.6}$ & $0.52^{+0.15}_{-0.16}$ & $6.80^{+0.08}_{-0.08}$ & $9.6^{+2.5}_{-3.7}$ & $2.78^{+0.70}_{-0.08}$ & 7.0/36 & $4.4\times10^{-8}$ \\ [1ex]
rms & & & $1.79^{+0.03}_{-0.05}$ & &  $0.07^{+0.21}_{-0.07}$ & & & & 17.8/54 & \\ [1ex]

30188-06-04-00 & $0.53^{+0.19}_{-0.53}$ & $4100^{+1200}_{-2000}$ & $2.03^{+0.03}_{-0.03}$ & $9.9^{+1.7}_{-0.7}$ & $0.54^{+0.17}_{-0.15}$ & $6.82^{+0.09}_{-0.09}$ & $9.6^{+2.9}_{-3.9}$ & $2.79^{+0.68}_{-0.05}$ & 6.1/36 & $5.1\times10^{-8}$  \\ [1ex]
rms & & & $1.90^{+0.03}_{-0.03}$ & & $0.33^{+0.12}_{-0.11}$ & & & & 38.9/54 & \\ [1ex]

30188-06-05-00 & $0.58^{+0.08}_{-0.58}$ & $8400^{+880}_{-2800}$ & $2.19^{+0.07}_{-0.04}$ & $11.6^{+3.3}_{-1.4}$ & $0.54^{+0.20}_{-0.17}$ & $6.81^{+0.10}_{-0.11}$ & $9.0^{+3.9}_{-2.9}$ & $2.81^{+0.18}_{-0.22}$& 6.0/36 & $6.0\times10^{-8}$ \\ [1ex]
rms & & & $2.01^{+0.03}_{-0.03}$ & & $0.10^{+0.10}_{-0.10}$ & & & &  28.2/54 & \\ [1ex]

30188-06-06-00 & $0.65^{+0.08}_{-0.45}$ & $9100^{+540}_{-1000}$ & $2.33^{+0.06}_{-0.06}$ & $11.5^{+4.5}_{-1.5}$ & $0.70^{+0.19}_{-0.20}$ & $6.80^{+0.08}_{-0.09}$ & $11^{+2.7}_{-3.7}$ & $2.75^{+0.17}_{-0.16}$ & 5.6/36 & $5.4\times10^{-8}$\\ [1ex]
rms  & & & $2.07^{+0.02}_{-0.04}$ & & $0.30^{+0.25}_{-0.13}$ & & & & 38.7/54 & \\ [1ex]

30188-06-07-00 & $0.70^{+0.07}_{-0.51}$ & $6600^{+350}_{-560}$ & $2.27^{+0.08}_{-0.06}$ & $10.4^{+2.7}_{-1.0}$ & $0.64^{+0.15}_{-0.20}$ & $6.81^{+0.08}_{-0.08}$ & $11^{+3.0}_{-3.4}$ & $2.78^{+0.23}_{-0.22}$ & 6.3/36 & $6.4\times10^{-8}$ \\ [1ex]
rms & & & $2.11^{+0.03}_{-0.03}$ & & $0.29^{+0.08}_{-0.11}$ & & & & 35.8/54 & \\ [1ex]

30188-06-08-00 & $0.70^{+0.07}_{-0.62}$ & $6500^{+310}_{-700}$ & $2.26^{+0.07}_{-0.15}$ & $10.3^{+3.4}_{-0.84}$ & $0.61^{+0.20}_{-0.16}$ & $6.82^{+0.09}_{-0.08}$ & $11^{+3.7}_{-3.8}$ & $2.83^{+0.60}_{-0.13}$ & 5.4/36 & $6.4\times10^{-8}$ \\ [1ex]
rms & & & $2.08^{+0.03}_{-0.03}$ & & $0.16^{+0.09}_{-0.09}$ & & & & 47.0/54 & \\ [1ex]

30188-06-09-00 & $0.66^{+0.08}_{-0.45}$ & $9600^{+480}_{-900}$ & $2.39^{+0.08}_{-0.05}$ & $12.8^{+5.3}_{-1.9}$ & $0.67^{+0.16}_{-0.20}$ & $6.79^{+0.08}_{-0.08}$ & $9.5^{+3.0}_{-3.1}$ & $2.73^{+0.14}_{-0.16}$ & 6.4/36 & $6.7\times10^{-8}$  \\ [1ex]
rms & & & $2.08^{+0.03}_{-0.03}$ & & $0.22^{+0.13}_{-0.12}$ & & & & 21.8/54 & \\ [1ex]

30188-06-10-00 & $0.52^{+0.12}_{-0.46}$ & $17000^{+3500}_{-4400}$ & $2.25^{+0.05}_{-0.04}$ & $11.8^{+4.1}_{-2.1}$ & $0.54^{+0.27}_{-0.14}$ & $6.82^{+0.11}_{-0.12}$ & $8.1^{+4.1}_{-4.0}$ & $2.80^{+0.17}_{-0.10}$ & 8.2/36 &$6.2\times10^{-8}$  \\ [1ex]
rms & & & $2.09^{+0.04}_{-0.04}$ & & $0.33^{+0.15}_{-0.13}$ & & & &  34.3/54  & \\ [1ex]

30188-06-11-00 & $0.70^{+0.06}_{-0.33}$ & $8600^{+380}_{-750}$ & $2.40^{+0.08}_{-0.07}$ & $11.9^{+4.1}_{-2.1}$ & $0.65^{+0.24}_{-0.17}$ & $6.79^{+0.09}_{-0.09}$ & $11^{+3.4}_{-3.6}$ & $2.78^{+0.22}_{-0.18}$  & 3.6/36 & $7.3\times10^{-8}$ \\ [1ex]
rms & & & $2.15^{+0.03}_{-0.06}$ & & $0.16^{+0.22}_{-0.08}$ & & & & 24.7/54 & \\
\end{tabular}
\caption{Fit results for the model \textsc{tbabs$\times$gabs$\times$(diskbb+nthcomp+nthcomp$\times$rfxconv)}. Errors indicate 90\% confidence intervals.}
\label{datatable}
\end{table*}

\begin{figure*}
\includegraphics[width=17cm,angle=0]{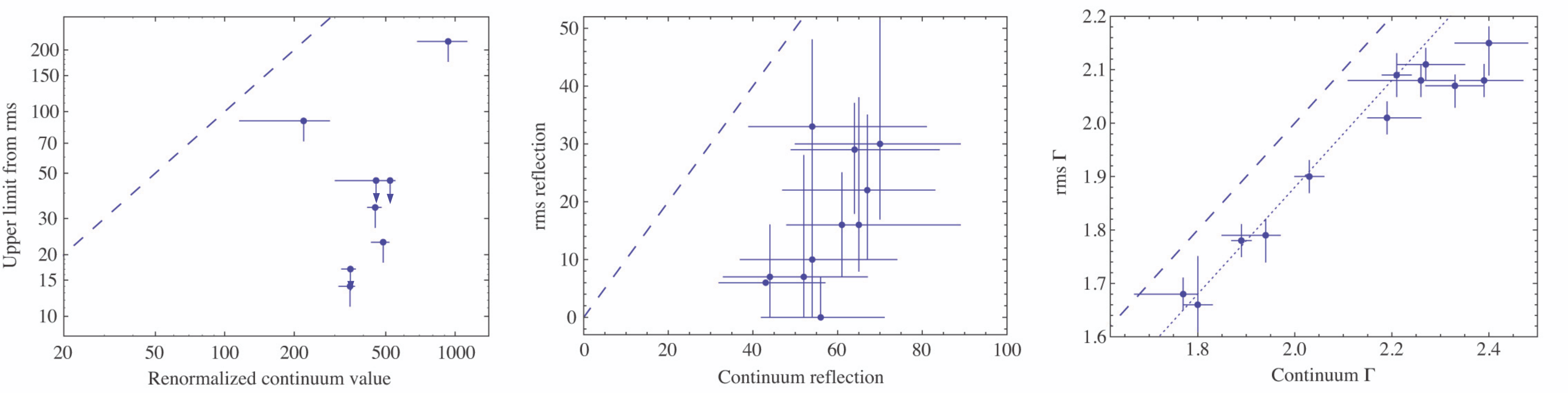}
\caption{Comparison of disc blackbody (left panel), amount of reflection (middle panel) and spectral index $\Gamma$ (right panel) between the continuum and variability
spectra. In all panels, the dashed line shows where the two are equal. As the rms spectra do not require any blackbody component, the left panel shows upper limits plotted against the strength of the blackbody that would be expected if simply scaling down the continuum spectrum to the level of the rms spectrum.}
\label{parameters}
\end{figure*}

The most obvious difference between the continuum and the spectrum of
the fast variability is in the strength of the blackbody and the
reflection component. While the continuum spectra in all but the
hardest states require a rather strong blackbody component, the
data require that the normalization of this
component is set to zero in the variability spectrum. The same is true
for Compton reflection, which varies between 0.4 and 0.7 for the
continuum, but is lower than 0.3 for all the spectra of the fast
variability and in the cases with harder spectra consistent with zero
reflection. (The ionization level of the reflector cannot be
constrained for the low levels of reflection in the fast variability
spectrum.) There are thus no signs of a disc in the spectrum of the
fast variability, either from its direct or reflected emission. 

Further, we note that $\Gamma$ in all cases is harder in the rms
spectrum than in the corresponding continuum spectrum. This shows that
the fast variability has its origin in a separate Comptonization
region where the plasma is more photon starved than for the overall
Comptonization in the source. We note that we keep the electron
temperature of the fast variability component frozen to the value for
the continuum for simplicity since the fast variability spectrum is
even more poorly constrained at higher energies than the continuum PCA
data. To examine possible differences in the electron temperature or
distribution between the origin of the fast varaibility and the
continuum, both higher energy data and a more physical spectral model
is required. However, our results already clearly show that the
Comptonization region itself is inhomogenous. The rms spectra enable
us to separate out the region where the fastest variability arises,
and this region clearly sees less seed photons than the flow as a
whole to produce its harder spectrum. 

The differences between the rms and continuum spectra are highlighted
in Fig.~\ref{parameters}. The figure contrasts the behaviour of the
disc blackbody, reflection and spectral index of the two spectra. As
the rms spectra do not require any blackbody component, the upper
limits are plotted against the strength of the blackbody as would be
expected if simply scaling down the continuum spectrum (left panel of
Fig.~\ref{parameters}). When the continuum spectrum is hard, the
disc blackbody is not significantly detected in the time averaged
spectrum so we do not plot these points as they give no constraints. 
However, in the softer states when the
disc component becomes more dominant, the upper limits from the rms
spectra clearly rule out any correspondingly strong blackbody in the
rms spectra. The two other panels in Fig.~\ref{parameters} directly
compare the strength of reflection (middle panel) and spectral index
(right panel) derived from the fits.

\begin{figure}
\includegraphics[width=12cm,angle=270]{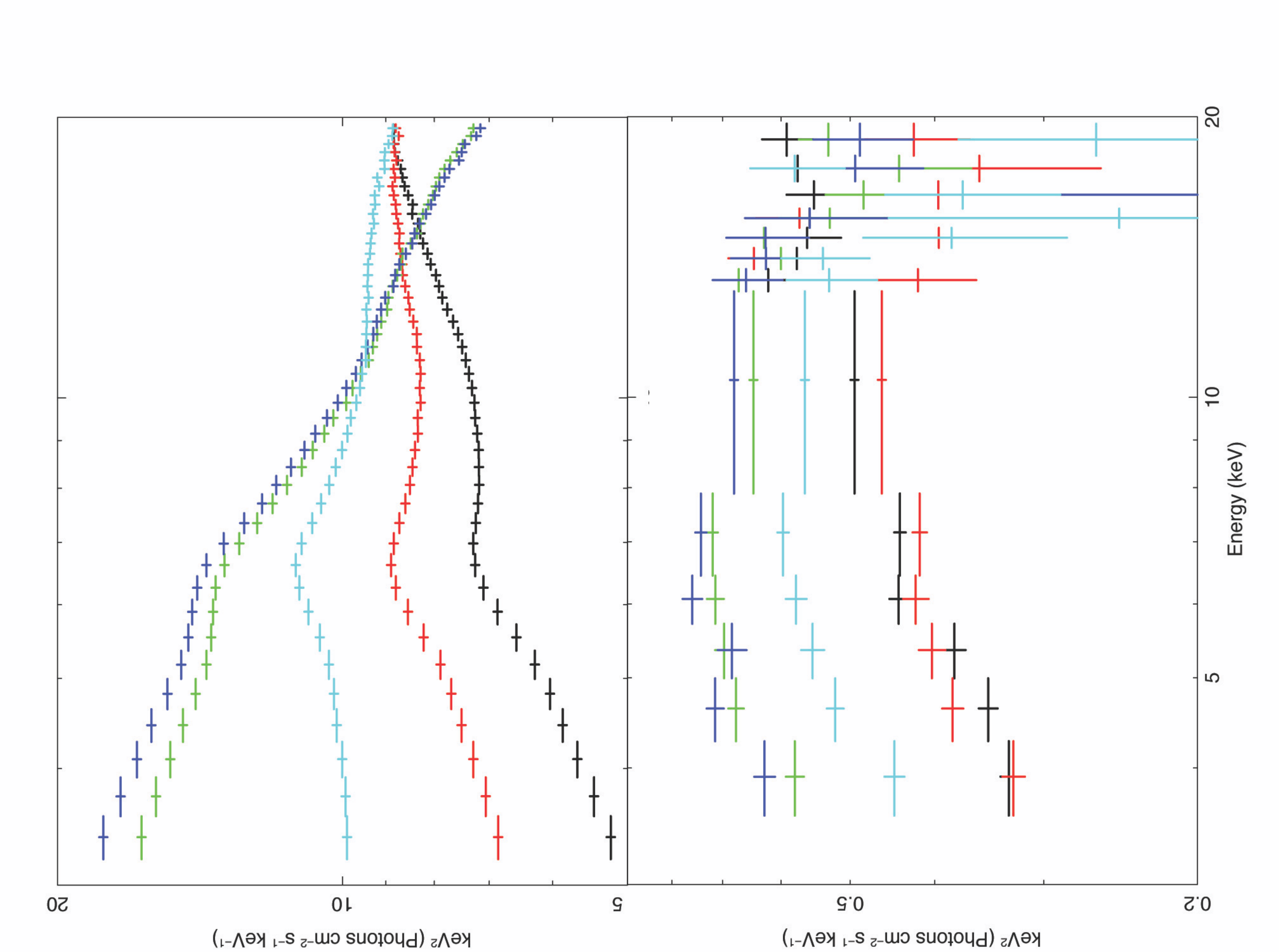}
\caption{Sample of PCA (upper panel) and rms (lower panel) spectra used in this study. Although the PCA spectrum changes considerably, the spectrum 
of the rapid variability show much less alteration.}
\label{specchanges}
\end{figure}

Another striking result from our analysis is the apparent stability of the variability spectrum, despite the fact that the source changes state as evidenced by the large differences in the Standard2 spectra. While the Standard2 spectra change considerably with the spectral index $\Gamma$ varying between 1.77 and 2.40, the shape of the variability spectrum varies between 1.66 and 2.15. This is clearly illustrated in Fig.~\ref{specchanges} where we show Standard2 spectra from the whole time interval (upper panel) and the corresponding variability spectra (lower panel). The rms spectrum is thus less sensitive to a changing accretion rate and to a changing influx of soft photons.

\subsection{Relation between fast variability and QPO}

\begin{figure*}
\includegraphics[width=17cm,angle=0]{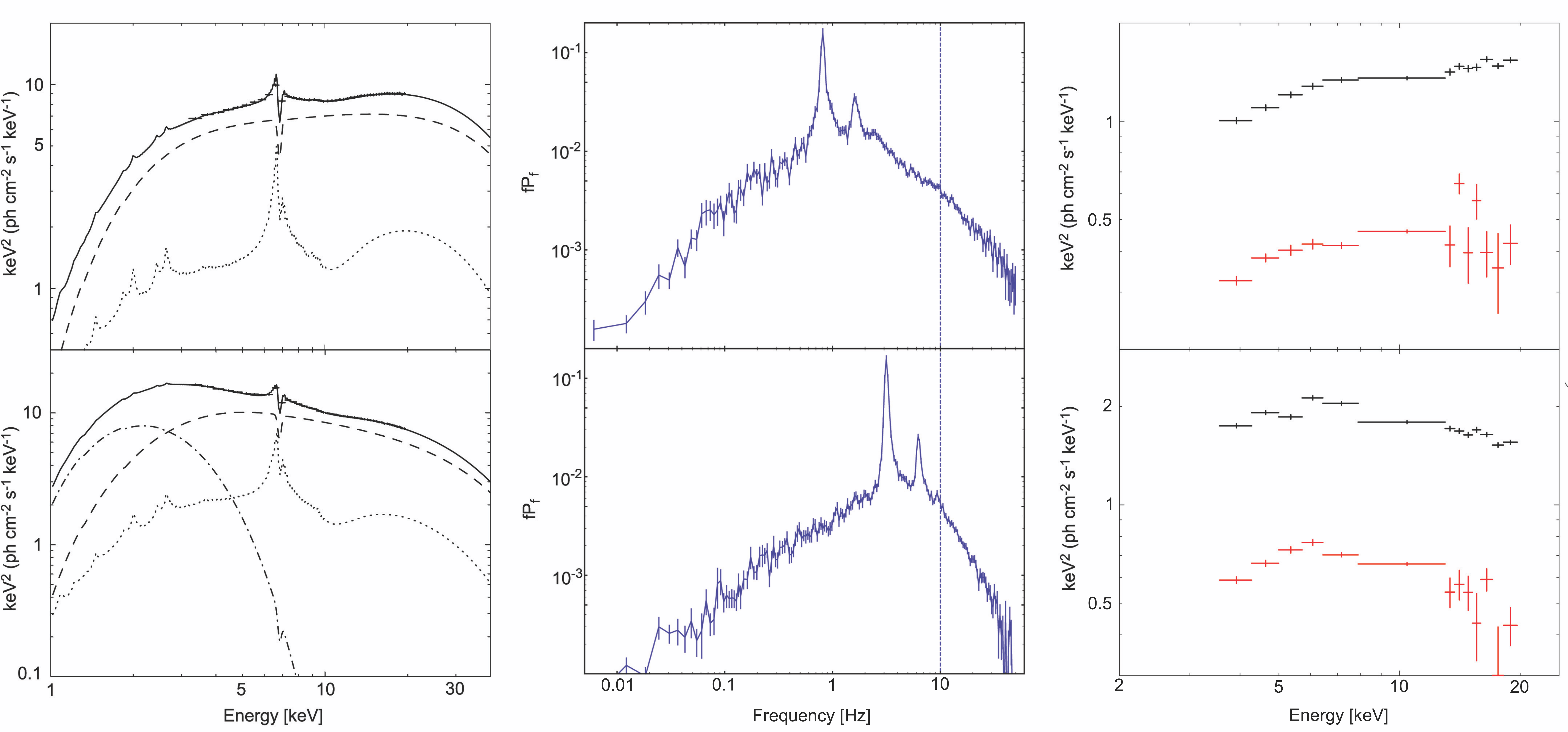}
\caption{Fit of PCA spectra to our model (left panels), the PDS of the observation (middle panels) and spectra of rapid variability and
QPO (right panels) for the observations 30188-06-01-02 (upper row) and 30188-06-08-00 (bottom row). The vertical line in the middle
panels show the 10 Hz lower boundary for the rapid variability spectra. }
\label{spectra}
\end{figure*}

Previous studies have looked at the spectrum of the QPO in both
XTE~J1550-564 and other sources \citep{sob06}. In order to compare the
rms spectrum to that of the QPO, we have extracted the spectrum of the
latter as described in Sect.~\ref{fourspec}. In Fig.~\ref{spectra}
we show a spectral fit to the PCA spectrum (left column), the PDS from
the observation (middle column), and the energy spectrum of the rapid
variability (right column). The two rows correspond to ObsIDs 30188-06-01-02 
(top) and 30188-06-08-00 (bottom), respectively. The right
column also shows the spectrum of the QPO.

It is clear that the QPO spectrum is quite similar to that of the
rapid variability (see also the neutron star spectra in 
Revnivtsev et al 2006). This would be expected if both originate in a
common physical region. There is perhaps a hint that the variability
spectra turn down more sharply towards lower energies, but our data do
not allow us to test whether this is a real effect. We note that the
close similarity between the spectra ensures that any possible
contribution from the QPO, despite our conservative limit of 10 Hz,
does not affect the shape of the rapid variability spectra.

The rms spectra of the QPO were modelled in detail by Sobolewska \&
{\.Z}ycki (2006). In contrast to the approach here, they used reflection
models to fit the shape of these spectra. However, our more
sophisticated reflection models clearly show that there is very little
reflection in the fast variability spectra, hence it seems likely that
the QPO rms spectra are similarly from Comptonized emission rather
than reflection. 

\section{Discussion}

It has long been clear that the accretion flow is inhomogeneous, with
components from a disc and corona. However, our results point to
inhomogeneity also within the coronal emission itself, with different
Comptonization spectra produced in different regions of the flow
\citep[see also][]{rev99,vel12}.

This is particularly evident in the softer observations. The time
averaged spectra clearly show strong reflection and a strong disc
component in addition to the Comptonization continuum. The fast
variability spectrum lacks the reflection and direct disc emission,
showing that the variability arises in a region of the corona where
the disc subtends a small solid angle.  However, the fast variability
spectrum is not simply that of the time averaged Comptonization
spectrum, as would be expected if the corona was homogeneous. Instead,
the fast variablity spectrum is harder than the time averaged
Comptonization continuum.  The variable part of the spectrum thus does
not represent the whole Comptonized region, only a part of it, and the
Comptonized component we see in the time averaged spectrum cannot
represent one uniform region but rather a superposition of at least
two regions (one variable at high frequencies and one not). The different
hardness and absence of reflection points to the variable part being
physically separated from the other Comptonized region/regions (see
also Yamada et al. 2013).

\begin{figure}
\includegraphics[width=8cm,angle=0]{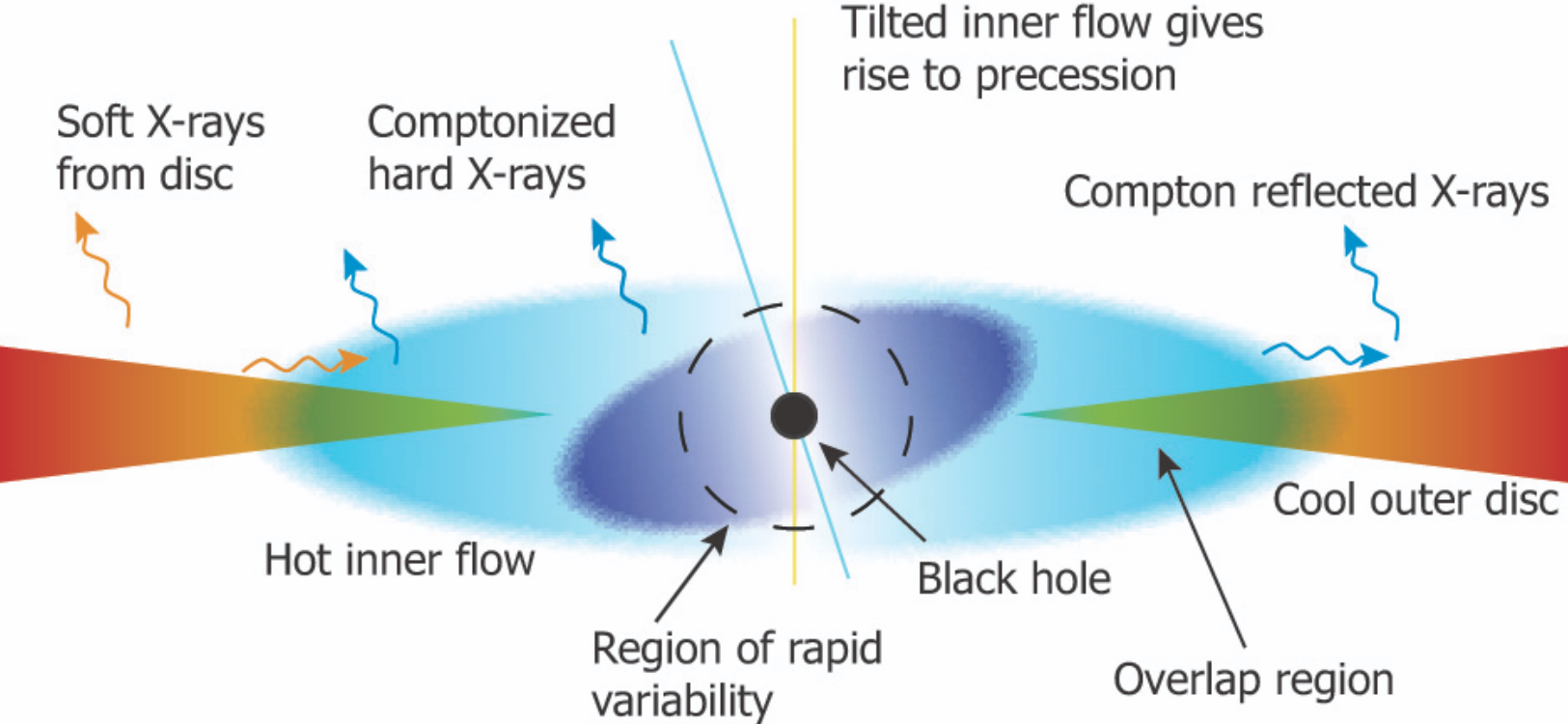}
\caption{Sketch of accretion geometry with different emission regions indicated. Our results indicate indicate that the region of rapid variability
sees very few disc photons, indicating that it is far away from the disc and close to the black hole.}
\label{geometry}
\end{figure}

This has an obvious geometric interpretation in the truncated disc/hot
inner flow picture where the inner region of the flow intercepts fewer
seed photons from the disc (thus has a harder spectrum) than the outer
regions which are closer, or even over the disc. In
Fig.~\ref{geometry} we sketch a picture of our preferred geometry and
origin of the fast variability (see also Yamada et al. 2013). This
framework also explains the observed hard time lags seen in black hole
binary systems \citep[see, e.g.,][]{miy89}.  If the variability arises
in the outer (softer) regions of the flow, and propagates inwards
where the hard emission is produced, such lags are naturally produced
\citep{kot01,are06}. We note that these lags will introduce
decoherence, and so suppress variability on the timescale of the lag
(Focke, Wai \& Swank 2005). However, \citet{cui00} show that the time
lags in XTE~J1550-564 are below a few milliseconds at frequencies
above 10 Hz so this effect will be small.

The differences between continuum and fast variability rms spectra are
not as great in the harder observations. However, better
signal-to-noise observations of Cyg X-1 reveals similar behaviour in
the hard states also \citep{rev99}. Since the time lags are
also present in these states \citep[see, e.g.,][]{miy89,tor11} then the flow
geometry is likely inhomogeneous here as well.

In the geometry of Fig.~\ref{geometry}, the rapid variability is
connected to a separate component, giving a straightforward
interpretation of our results. However, this may not be unique. There
is an alternative model involving spectral evolution of a single,
homogeneous source (a magnetic flare whose Comptonized spectrum
evolves from soft to hard as the flare rises above the disc) which
also produces the observed time lag behaviour \citet{pf99}. Clearly in
this scenario the fast variability spectrum does not imply a separate
component.  However, we note that this specific model does not appear
to match all aspects of the data. While they do not explicitly show an
rms spectrum, it is clear that the model leads to spectral pivoting
around 10-15~keV for their model parameters (a hard state), so there
should be a minimum in the variability spectrum at this energy \citep[as
also shown by the power spectrum at 27~keV having less high frequency
variability that that at 3~keV: figure 2a of][]{pf99}. By contrast
our data show that this is not the case, rather, we see a peak in
variability in this energy band.

\subsection{Spectral evolution}

Our interpretation of an inhomogeneous geometry is also consistent
with the observed long term spectral evolution seen in the data. 
The source starts in the low/hard state, where the disc is truncated
further from the black hole, then the spectra get softer and the
direct disc and reflection components get stronger as the inner radius
of the truncated disc moves in towards the black hole as the mass
accretion rate increases. 

The variability spectra also show some spectral evolution, but much
less than seen in the time averaged spectra. The fast variability
spectrum is thus less sensitive to the increase in seed photons than
is the total spectrum. Assuming a spherical geometry for the hot inner
flow, purely geometrical considerations reveal that the fraction of
disc photons intercepted by the flow interior to the disc truncation
radius does not depend on the radius of the sphere (as long as it
tracks the evolution of the inner disc radius). Therefore, even if the
total spectrum will change dramatically in response to variations in
accretion geometry, the energy spectrum of the rapid variability - if
connected to the hot innermost flow - will retain a more or less
constant ratio between the power supplied to the electrons and the
power in the soft seed photons throughout the evolution.

\subsection{Connection to the QPO}

Recently \citet{ing11,ing12} presented the first physical model for
creating broad band variability and QPOs in the framework of the disc
truncation scenario. They showed that propagating mass accretion rate
fluctuations in the hot flow can match the broad band power spectral
properties seen.  They also model the QPO as a vertical precession of
the same inner part of the hot inner flow (see Fig.~\ref{geometry}). The 
fact that the QPO and rapid variability have very similar spectra support 
this picture as a vertical precession can only occur for regions inside the 
disc truncation radius. Thus the precessing region can only be interior to 
the disc truncation radius, so is associated with the same radii as produce 
the fastest variability. This predicts that the QPO spectrum also has little 
contribution from the disc and reflected emission. We will test this
directly in a future paper (Axelsson et al. 2013).

\section{Summary and Conclusions}

We have modelled the energy spectrum of the fast variability (10-50
Hz) in the Galactic black hole binary XTE~J1550-564 and compared it to
the time averaged PCA continuum. We find that the spectrum of the
rapid variability differs from the time-averaged continuum in that it
requires no thermal disc component, is significantly harder and shows
much less or no Compton reflection. The spectrum of the fast
variability also shows less evolution of its spectral shape with
increasing accretion rate than the time averaged continuum. We
conclude that the rapid variability arises in a hot inner flow close
to the black hole, a region which does not intercept  a large amount of
seed photons from the disc. Thus it is not strongly affected by the increase
of soft seed photons following a change in the geometry of the
accretion flow in the transition from a hard to
a very high state. The region is likely to collapse once the source
enters a classical soft state, with the inner disc radius reaching all
the way in to the last stable orbit.

The fact that the spectrum of the fast variability differs from the
continuum even after removing the blackbody and reflection components
further suggests that the optically thin emission usually modelled as
one single Comptonization component is likely a superposition of
two or more components -- one variable at high frequencies
and one not. This shows the need for more complex models of the 
total spectra of black hole binaries in general.

\section*{Acknowledgments}
This work was supported by The Royal Swedish Academy of Sciences and The Swedish Research Council (VR 623-2009-691). LH acknowledges 
support from the Wenner-Gren Foundations. This research has made use of data obtained through the High Energy Astrophysics Science Archive 
Research Center (HEASARC) Online Service, provided by NASA/Goddard Space Flight Center.

\end{document}